\documentclass[twocolumn,aps,floatfix,superscriptaddress,showpacs]{revtex4-1}

\usepackage{amsmath}
\usepackage{dsfont}
\usepackage{amssymb}
\usepackage{graphicx}

\renewcommand{\>}{\rangle}
\newcommand{\<}{\langle}

\def\XXint#1#2#3{{\setbox0=\hbox{$#1{#2#3}{\int}$}
     \vcenter{\hbox{$#2#3$}}\kern-.5\wd0}}

\begin{document}

\title{Quantum ripples over a semi-classical shock}
\author{Eldad Bettelheim}
\affiliation{Racah Institute of Physics, Hebrew University, Jerusalem 91904, Israel}
\author{Leonid Glazman}
\affiliation{Department of Physics, Yale University, New Haven, CT 06520, USA}

\begin{abstract}
The evolution of an initially smooth spatial inhomogeneity in the density of a one-dimensional Fermi gas is well described by classical mechanics. The classical evolution  leads to the formation of a shock wave: the density develops kinks in its coordinate dependence. We show that quantum corrections to the shock wave produce density ripples which run off the kinks. Despite their quantum origin, the amplitude and period of the ripples are expressed only in terms of  classical objects derived from a smooth density profile.
\end{abstract}

\maketitle


In one-dimensional many-body physics, numerous long-wavelength properties of a quantum system are faithfully represented by the dynamical properties of a continuous liquid. Such representation is at the heart of a powerful bosonization method \cite{Giamarchi:Bosonization:Book,Haldane:Luttinger:Liquid,Glazman:Imambekov:Luttinger:Review} . Small-amplitude perturbations of the liquid's density may be considered in a harmonic approximation. In that approximation, waves of density are linear, and there is no difference between the quantum and the classical dynamics of the density perturbations. At larger amplitudes, waves become nonlinear; it is not clear then, if the classical and quantum dynamics remain indistinguishable even in the long-wavelength limit. 

In a liquid made of fermions, higher density is associated with higher Fermi momentum and higher particles velocities. Described classically, a higher-density solitary segment moves faster than its lower-density periphery. That leads to a wave overturn and the formation of shocks in the density $\rho(x,t)$, {\it i.e.}, points $x_\gamma (t)$ where the $x$-dependence of $\rho$ looses its analyticity. Formation of shocks can be seen within a classical continuous medium description in terms of the Riemann-Hopf equation   \cite{Landau:Book:Fluid:Mechanics,Whitham:Book,Riemann:Hopf:Original:Riemann}. It is insensitive to the interaction strength, which can be set to zero (the free fermions case). 
The classical description of propagation of shock waves  in a  gas of quantum particles is widely accepted across the fields of quantum physics from string theory \cite{Polchinksi:Fermi:Sea} to cold atomic gases \cite{Bulgac:Shock:Waves} . In case of free fermions, however, one may
 also investigate the fully-quantum evolution of the many-body wave function, which is a Slater determinant of free-propagating single-particle states. 

We consider the shock formation in the semiclassical limit of the quantum evolution of free fermions. Our main finding is the appearance of oscillatory structure in $\partial\rho/\partial x$ accompanying each of the shock points $x_\gamma (t)$. This structure, quantum in origin, has a characteristic scale of variation fully determined by the solution of the classical problem. (In that respect, the free-particle many-body wave function associated with the shock formation bears some resemblance to the WKB wave function of a single particle in an external potential.)  In classical physics, the formation of shocks   \cite{Riemann:Hopf:Original:Riemann}\footnote{See also \S 101 of  \cite{Landau:Book:Fluid:Mechanics}} 
is associated with the  presence of  velocity gradients. The inverse time to develop a shock can be estimated as $t_S\sim |dv/dx|$, where $v(x)$ is the initial velocity distribution. If the velocity gradient comes from a ``bump'' $\Delta\rho$ in the density of fermions, then the corresponding density gradient, $d\rho/dx=(m/\hbar)dv/dx$ becomes large in the semiclassical limit, $\hbar\to 0$ at fixed $dv/dx$ (hereinafter $m$ is the mass of a fermion). Together with $|d\rho/dx|$, the total number of particles $\Delta N$ in the density bump scales as $1/\hbar$. We show that $1/\Delta N$ can be consistently used as a small parameter in the semiclassical expansion. The characteristic length of the oscillatory structure around a shock point scales as $(\Delta N)^{-2/3}$.

We are considering the time evolution of a solitary density maximum $\rho(x,t)$, which initially (at $t=0$) has spatial extent $\Delta x$ and amplitude $\Delta\rho$,
\begin{align}\label{Scales}
\rho(x,0) = n_0 + \Delta \rho f\left(\frac{x}{\Delta x} \right)\,.
\end{align}
Here the dimensionless positive function $f(s)$ reaches a maximum, $f(0)\sim 1,$ at $s=0$, has spatial extent $|s|\sim 1$, and decays outside that region, $f(s\to\pm\infty)=0$, revealing the background equilibrium density $n_0=K_F/\pi$ of a Fermi sea ($|k|\leq K_F$) of spinless particles. We aim to describe the motion of {\it semiclassical} perturbations. That leads us to assume that many particles go into the creation of the perturbation, $\Delta N = \Delta \rho \Delta x \gg 1$.  Our further assumption is that the density perturbation is small compared to the equilibrium density, $\Delta \rho \ll n_0$. The two latter assumptions mean that $K_F\Delta x\gg \Delta N\gg 1$; therefore, the density perturbation consists of particles and holes residing in the narrow vicinities $\delta k \sim K_F \Delta \rho/n_0$  of the Fermi points $\pm K_F$. 

There are several characteristic time scales for the evolution of the density perturbation. First, it will decompose into right- and left-moving modes, associated with particle-hole excitations around each one of the Fermi points. The characteristic time associated with this process is $t_{LR} =(m/\hbar)(\Delta x/n_0)$. Next, there is a characteristic time over which the wave packet representing a single fermion spreads over a spatial scale $\Delta x$ of the initial many-body state. This time is given by $t_Q =(m/\hbar)(\Delta x)^2.$  On this time scale the evolution ceases to be semiclassical.  Finally, in the course of the semiclassical evolution of the density perturbation, it may develop a shock, in analogy to the one-dimensional classical hydrodynamics. This effect has been discussed, {\it e.g.}, in \cite{Bettelheim:Abanov:Wiegman:QH:Edges}. The corresponding characteristic time scale is $t_{S} = (m/\hbar)(\Delta x/\Delta \rho)$. 

Our assumptions $\Delta N \gg 1$ and $\Delta \rho \ll n_0$ lead to the time scales hierarchy, $t_{LR}\ll t_S\ll t_Q$ . Namely,  first the perturbation splits into left and right moving modes, then classical shocks have a chance to appear in each of the modes, and at much later times the semiclassical nature of the evolution breaks down.  We focus on the evolution occurring on time scale $\sim t_S$.

Since we are interested in the semiclassical regime, it is convenient to use the Wigner function,
\begin{align}\label{WignerDefintion}
W(x,k,t) =  \int \left\<\Psi|
 \psi^\dagger\left(x +\frac {y}{2},t\right) \psi \left(x - \frac {y}{2} ,t \right) |\Psi\right\> e^{-iky }dy,
\end{align}
where $|\Psi\>$ is the initial ($t=0$) state of the system.

In equilibrium, $|\Psi\>$ is a state where all single particle levels with momenta between $-\hbar K_F$ and $\hbar K_F$ are occupied, while all other states are empty. Here $K_F = \pi n_0$. This state leads to a Wigner function given by
$
W(x,k,t) = \theta \left(K_{F}- k \right) \theta\left(k +K_{F}\right).
$ The Wigner function is interpreted as a distribution in phase space of the Fermions, which, in equilibrium, occupy the region $-K_F < k <K_F$ for all $x$, namely a band in phase space.

We can think about a semiclassical state in which a single mode (either right- or left-moving) of density perturbation is excited, as a state in which one of the Fermi wave numbers ($-K_F$ or $K_F$) is allowed to vary with the space-time coordinates. From now on, and without loss of generality, we shall assume only a right moving mode, in which case the Fermi number $K_F$ becomes time and space dependent, to be described by a function $k_F(x,t)$, while $-K_F$ stays fixed. This heuristic picture translates into the following {\it ansatz} for the Wigner function:
\begin{align}\label{SimpleAnsatz}
W(x,k,t) = \theta(k_F(x,t) - k)\theta(k+K_F).
\end{align}
We will see that Eq.~(\ref{SimpleAnsatz}) is a good approximation to the Wigner function everywhere except the vicinities of the shock points.
Knowledge of the Wigner function allows one to compute the density, as these two are related as follows:
\begin{align}\label{DensityThroughW}
 \rho(x,t) = \int W(x,k,t)\frac{dk}{2\pi}.
\end{align}
Given initial conditions for the Wigner function, one may compute it at later times making use of the evolution equation,
\begin{align}\label{Wevolution}
\partial_t W(x,k,t) = \frac{\hbar k}{m} \partial_x W(x,k,t).
\end{align}
The proof of Eq.~(\ref{Wevolution}) consists of the application of the Schr\"{o}dinger equation to the fermions in Eq.~(\ref{WignerDefintion}) and integration by parts. The initial value problem for the evolution equation may be solved by the method of characteristics, which yields
\begin{align}\label{Wsolution}
W(x,k,t) = W\left(x-\frac{\hbar k}{m}t , k,0\right).
\end{align}
Applying Eq.~(\ref{Wsolution}) to ansatz (\ref{SimpleAnsatz}) one sees that $k_F(x,t)$ must satisfy:
\begin{align}\label{pFsolution}
k_F(x,t) = k_F\left(x-\frac{\hbar}{m} k_F(x,t) t,0 \right).
\end{align}
This  equation determines $k_F(x,t) $ implicitly, given the initial conditions $k_F(x,0)$.

The relation to the classical shock wave physics in one dimensional hydrodynamics may be realized by excluding $\hbar$ from Eq.~(\ref{pFsolution}) with the transformation $v_F(x,t)=(\hbar/m)k_F(x,t)$
and noticing that $v_F(x,t)$ is in fact a  solution to
\begin{align}\label{Hopfcomoving}
\partial_t v_F(x,t) + v_F(x,t)  \partial_x  v_F(x,t) =0.
\end{align}
This Riemann (or Riemann-Hopf) equation \cite{Riemann:Hopf:Original:Riemann} is well known to give rise to shock waves.

\begin{figure}[t!!!]
\begin{center}
\includegraphics[width=8.5cm]{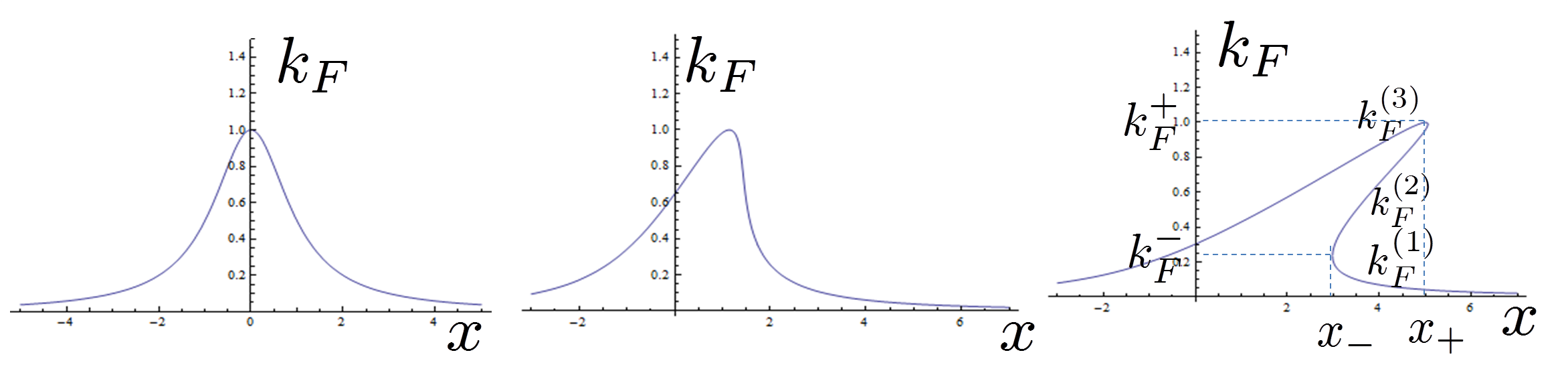}
\caption{ Function $k_F(x,t)=mv_F(x,t)/\hbar$ at initial time (left), evolved according to Eq.~(\ref{Hopfcomoving}) at an intermediate time (middle), and at a late time (right) where it becomes multi-valued. 
 \label{overhang}}
 \end{center}
\end{figure}

In the following we make use of the inverse function, $x_F(k,t)$, defined as the solution to the following equation:
\begin{align}\label{xFdefinition}
k_F(x_{F},t) = k.
\end{align}
The evolution of $k_F(x,t)$ determines the evolution of $x_F(x,t)$ as follows:
\begin{align}\label{xFevolution}
x_F(k,t) = x_F(k,0) + \frac{\hbar}{m}kt.
\end{align}
Therefore, the derivative $\partial x_F/\partial k = x'_F(k,0) + \frac{\hbar}{m} t$. If $x_F'(k,0)$ is negative, there is always a positive time at which $\partial x_F(k,t)/\partial k=0$. This equation, together with Eq.~(\ref{xFevolution}) determines the time evolution of points where
$|\partial k_F(x,t)/\partial x|=\infty$. This divergence is a manifestation of the shock phenomenon; it is also termed as the `gradient catastrophe'.  After the first time the function $k_F(x,t)$ displays an infinite derivative, the solution to Eq.~(\ref{pFsolution}) becomes multi-valued as shown in Fig.~\ref{overhang}. Without loss of generality, we assume that  the multi-valued region is bounded by two points, $x_-(t)$ and $x_+(t)$, between which the function $k_F(x,t)$ has three branches, enumerated by $k_F^{(1)}(x,t)< k_F^{(2)}(x,t)<k_F^{(3)}(x,t).$

Regardless of the fact that $k_F(x,t)$ is multi-valued,  initial conditions such as (\ref{SimpleAnsatz}) lead to a Wigner function which is $1$  in a region of phase space between the curve $k_F(x,t)$ and the horizontal line $k= -K_F$, while the 
\begin{figure}[t!!!]
\begin{center}
\includegraphics[width=7cm]{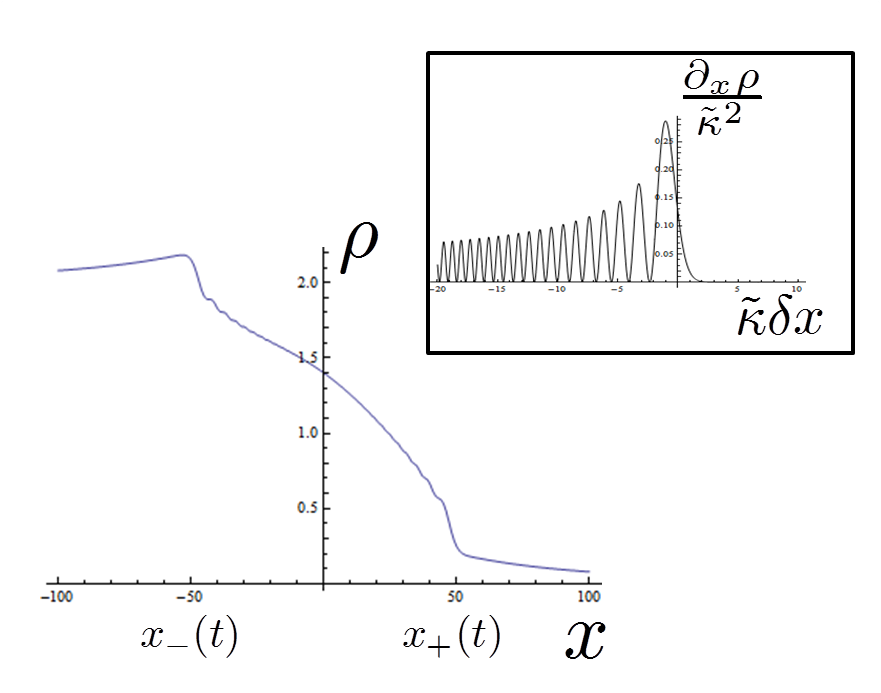}
\caption{A typical density profile. The classical trailing and leading shock points are $x_-(t)$ and $x_+(t)$, respectively; cf. Eq.~(\ref{rho0Near}). The quantum ripple effect is amplified in the derivative $\partial_x\rho (x,t)$. Inset:
${\partial_x \rho}/{\tilde{\kappa}^2}$ as a function of $\delta x = x-x_+(t)$ in the region next to $x_+(t)$, see Eq.~(\ref{deltaXrhoIsAi2}).
 \label{typicalRho}}
 \end{center}
\end{figure}
Wigner function is $0$ outside this region. Here the curve of $k_F(x,t)$ is understood to contain all the branches of the function, just as the curve described in the right panel of Fig. \ref{overhang}.

The integral in Eq.~(\ref{DensityThroughW}) may easily be performed to obtain the density. The result is:
\begin{align}
&2 \pi \rho^{cl}(x,t) = K_{F}+k_F^{(3)}(x,t)\! -\! k_F^{(2)}(x,t)\! +\! k_F^{(1)}(x,t),\,\,\, x\in I \nonumber \\
&2 \pi \rho^{cl}(x,t) =K_F + k_F(x,t),\quad x\notin I \label{rho0},
\end{align}
where $I=[x_-(t), x_+(t)]$ denotes the spatial segment of multi-valued behavior of $k_F$ . The superscript $cl$ in $\rho^{cl}$ denotes that (\ref{rho0}) gives only the result of classical theory, on which we want to improve in the vicinities of points $x=x_\pm$. 

Next to $x_-(t)$ and $x_+(t)$ two branches of the function $k_F(x,t)$ meet. Explicitly, $k_F^{(1)}$ and $k_F^{(2)}$ become equal to each other at $x_{-}(t)$:
\begin{align}
k_F^{(1)}(x_{-}(t),t) = k_F^{(2)}(x_{-}(t),t) \equiv k_F^-(t)\,,
\end{align}
and similarly at $x_{+}(t)$:
\begin{align}
k_F^{(2)}(x_{+}(t),t) = k_F^{(3)}(x_{+}(t),t) \equiv k_F^+(t)\,.
\end{align}
The functions $k_F^\pm(t)$ are solutions of the equation $\partial x_F(k,t)/\partial k=0$. The branches $k_F^{(1,2)}$  display a square-root  behavior next to $x_-(t)$:
\begin{align}\label{2root-}
k_F^{(2,1)}(x,t) - k_F^-(t) = \pm \alpha^-(t)\sqrt{x-x_-(t)}+\dots,
\end{align}
while $k_F^{(3)}$ is regular next to this point.  At the same time next to $x_+(t)$ we have:
\begin{align}\label{2root+}
k_F^{(3,2)}(x,t) - k_F^-(t) = \pm \alpha^+(t)\sqrt{x_+(t)-x}+\dots,
\end{align}
while now $k_F^{(1)}(x,t)$ is regular. The ellipsis denote
regular terms.

Near the branch meeting points, $\rho(x,t)$ takes the approximate form:
\begin{align}
\label{rho0Near}
\rho^{cl}(x,t) -  \rho^{cl}(x_\pm(t),t)  \simeq \pm \frac{ \alpha^\pm(t)}{\pi} \sqrt{\mp(x-x_{\pm}(t))} \,.
\end{align}
These square-root kinks (which may be observed in Fig.~\ref{typicalRho} next to $x_-(t)$ and $x_+(t), $ albeit rounded by  quantum corrections)  point to the non-analytical behavior of $\rho (x,t)$ at $x=x_\pm(t)$. The corresponding coefficients are determined by  initial conditions, $[\alpha^\pm(t)]^2=2(\partial^2_k x_F)^{-1}$. According to Eq.~(\ref{xFevolution}), the derivative $\partial^2_k x_F=d^2x_F(k,0)/dk^2$ at fixed $k$ is time-independent. To find $\alpha^\pm(t)$ one should evaluate the said derivative at $k=k_F^\pm (t)$.

Our goal is to evaluate quantum corrections to the classical density profile (\ref{rho0Near}). To that end we start with introducing a family of many-body coherent  states of free fermions \cite{Stone:book} with smooth density profiles. We are interested in times $t\gg t_{LR}$, so it is sufficient to include only, say, right-movers in the consideration. The corresponding fermion density operator is $\hat{\rho}^R = \psi^{(R) \dagger}(x) \psi^{(R)}(x)$, and the field operators contain only Fourier harmonics with $k>0$,
\begin{align}
\psi^{(R)}(x) = \int_{k>0}  \psi_{k} e^{ ikx} \frac{dk}{2\pi}\,.
\end{align}
The coherent states
\begin{align}\label{CoherentState}
|\Psi\rangle = e^{i \int  \Phi(x) \hat{\rho}^R (x) dx} | 0 \rangle ,
\end{align}
are parametrized by the function $\Phi(x)$ with a transparent meaning, $2\pi\rho^{cl}=K_F+2\pi\Phi'(x)$.

The standard bosonization methods~\cite{Stone:book}   yield
\begin{align}\label{Kernel}
\langle\Psi| \psi^\dagger(x_1) \psi(x_2) |\Psi\rangle =  \frac{e^{i (\Phi(x_1)- \Phi(x_2))}}{i2\pi(x_1-x_2)},
\end{align}
where we customarily dispensed with the additive $\Phi$-independent term which formally vanishes in the large-$K_F$ limit, $K_F|x_1-x_2|\to\infty$. 

To proceed with computing the Wigner function $W(x,k,0)$ we perform a gradient expansion in the exponent of Eq.~(\ref{Kernel}), to obtain:
\begin{align}
\<\Psi | \psi^\dagger\left(x+\frac{y}{2},0\right) \psi\left(x-\frac{y}{2},0\right) | \Psi \> \approx \frac{e^{i \left[\frac{k''_F(x )}{24} y^3+  k_F(x )y\right] }}{i2\pi y}.
\end{align}
The Fourier transform with respect to $y$ yields an approximation to the Wigner function \cite{Bettelheim:Wiegmann:Fermi:Number}:
\begin{align}\label{WignerAiry}
W(x,k,0)\approx {\rm Ai}_1\left ( 2^{2/3} \kappa(x,0) (k - k_F(x,0)) \right),
\end{align}
where
\begin{align}\label{Airy1Def}
{\rm Ai}_1 (x)=\int_s^\infty {\rm Ai}(s') ds'  = \int e^{i \left(\frac{q^3}{3} + q x\right)}\frac{dq}{2\pi i q},
\end{align}
and  $\kappa$ is the $t=0$ value of
\begin{align}\label{kappadef}
\kappa(x,t)=| \partial_x^2 k_F(x,t)/2|^{-1/3}\,.
\end{align}
The function ${\rm Ai}_1(s)$ changes monotonically from $1$ at $s\to-\infty$ to $0$ at $s\to\infty$ on the scale $|s|\sim 1$. Therefore at large $\kappa(x,0)$ the Wigner function of Eq.~(\ref{WignerAiry}) approaches its classical form for the right-movers,
\begin{align}
W(x,k,0) \approx \theta (k_F(x)-k).
\end{align}
This confirms the ansatz (\ref{SimpleAnsatz}), where the second factor $\theta(k+k_F)$ is now missing, as accounting exclusively for the right-movers formally corresponds to sending the left Fermi point, $-K_F$,  to $-\infty$.

Estimating $\kappa(x,0)$ with the help of Eq.~(\ref{Scales}) as $\kappa \sim \Delta \rho^{-1/3} \Delta x^{2/3}$, we see that  smearing of the Fermi step-function, as described by Eq.~(\ref{WignerAiry}), is smaller by  the parameter $(\Delta N)^{-2/3}$ than the shift $\sim\Delta N/\Delta x$ of the Fermi wave vector $k_F(x)$ from its equilibrium value. The same semiclassical parameter is needed to justify the gradient expansion which we used to derive Eq.~(\ref{WignerAiry}). To assess the accuracy of that approximation, we may consider the correction to Eq.~(\ref{WignerAiry})  resulting from the sub-leading term in the gradient expansion.  The correction has the form:
\begin{align}
\label{Correction}
\delta W
\sim\frac{\kappa^5 k^{''''}_F(x)}{2^{2/3}5!} {\rm Ai}'''' \left(2^{2/3} \kappa(x,0) (k - k_F(x,0))\right).
\end{align}
Now we again use Eq. (\ref{Scales}), to find $k''''_F \sim\Delta \rho \Delta x^{-4}$, in addition to the estimate for $\kappa(x,0)$. Substitution of these estimates in Eq.~(\ref{Correction}) yields $\delta W/W\sim\Delta N^{-2/3}$ at $|k-k_F(x)|\lesssim 1/\kappa(x,0)$.

We wish now to propagate the initial conditions (\ref{WignerAiry}) in time.
It will be more convenient, however, first to switch to a representation involving  only the inverse function, $x_F(k,t),$ defined in (\ref{xFdefinition}), rather than $k_F(x,t)$. We first focus on the factor $k-k_F(x,0)$ in (\ref{WignerAiry}), aiming to get rid of the function  $k_F$ and replace it by  $x_F$.  Making use of (\ref{xFdefinition}) to replace $k$ by $k_F(x_F(k,0),0)$ and Taylor expanding the function $k_F$ around $x_F(x,0)$ we find:
\begin{align}\label{Taylor}
k_F(x,0)-k = \frac{x-x_F(k,0)}{x_F'(k,0)}+ \dots,
\end{align}
where we have also used $k'_F(x_F(k,0),0) = 1/x_F'(k,0)$. We substitute the first term of expansion (\ref{Taylor})  into (\ref{WignerAiry}). This is justified at $x-x_F(k,0)\ll \Delta x$, and at the same time sufficient to allow the argument of the ${\rm Ai}_1$ function in (\ref{WignerAiry}) to vary in the range $|s|\sim (\Delta N)^{2/3}\gg 1$. In that range, the Wigner function closely approaches its limits $1$ and $0$ at negative and positive values of the argument, respectively.

We turn our focus now to $\kappa(x,0)$ appearing in (\ref{WignerAiry}). Since $\kappa,$ as defined in (\ref{kappadef}), contains $\partial_x^2 k_F(x,0)$, we need to replace it by an expression involving only $x_F(k,0)$. To this end we use the  following classical differential identity:
\begin{align}\label{ODEidentity}
\left.\partial_x^2k_F(x,0)\right|_{x=_F(k)}  = - \frac{x_F''(k,0)}{x_F'(k,0)^3}.
\end{align}
Combining Eqs.~(\ref{Taylor}) and (\ref{ODEidentity}) we may re-write Eq.~(\ref{WignerAiry}) as
\begin{align}\label{xW}
W(x,k,0) \approx {\rm Ai}_1 \left( 2^{2/3} \tilde{\kappa}(k) (x-x_F(k,0))\right),
\end{align}
where
\begin{align}\label{kappaDef}
\tilde{\kappa}(k)=\tilde{\kappa}(k,t) = \left|\partial_k^2x_F(k,t)/2 \right|^{-1/3}.
\end{align}
is independent of time, according to Eq.~(\ref{xFevolution}).

We are now ready to compute the Wigner function $W(x,k,t)$ and quantum corrections to density at later times. Combining Eqs.~(\ref{Wsolution}), (\ref{xFevolution}), and (\ref{xW}) we easily find:
\begin{align}\label{WignerAiryX}
W(x,k,t) \approx\ {\rm Ai}_1 \left( 2^{2/3} \tilde{\kappa}(k) (x-x_F(k,t))\right)\,.
\end{align}
Given the validity of Eq.~(\ref{WignerAiryX}) for all times, it remains only to integrate it over $k$ to obtain the density. Using the representation
(\ref{Airy1Def}) of ${\rm Ai}_1$,
one arrives at:
\begin{align}\label{deltaXthoFirstGo}
\partial_x \rho(x,t) \approx \int e^{ i \left[ \frac{|x''_F(k)|}{24} q^3 + q (x - x_F(k,t))\right]} \frac{dq}{2\pi}\frac{dk}{2\pi}.
\end{align}
Next we implement a gradient expansion of function $x_F(k,t)$ around a branch meeting point, $x_\pm(t)$. Using this expansion in Eq.~(\ref{deltaXthoFirstGo}) and changing variables there to $K = k+\frac{q}{2}$ and $Q=k-\frac{q}{2}$, we derive 
\begin{align}
&\partial_x \delta\rho^\pm(x,t) \label{KQ}\\
&\approx \int e^{i \left[\frac{|x''_F(k_{F}^\pm(t))|}{6}(K^3 - Q^3) + (x-x^\pm(t)) (K-Q) \right]} \frac{dQ}{2\pi}\frac{dK}{2\pi}. \nonumber
\end{align}
Note that here we are computing only the singular contribution $\delta \rho^\pm(x,t)$ next to the branch meeting points, 
\begin{align}
\delta \rho^{\pm}(x,t) = \pm \left ( \rho^{}(x,t) - \rho^{cl}(x_\pm(t),t)\right),
\end{align}
rather than full density $\rho(x,t)$. Integration over $Q$ and $P$ in Eq.~(\ref{KQ}) factorize, leading to the result:
\begin{align}\label{deltaXrhoIsAi2}
\partial_x \rho^\pm(x,t) \approx \tilde{\kappa}^\pm(t)^2 \left[ {\rm Ai}(\tilde {\kappa}^\pm(t) \delta x_\pm ) \right]^2\,.
\end{align}
Here  the distances, $\delta x_\pm = \mp (x- x_\pm(t)),$ are measured from the classical shock points, and inverse length scales $\tilde{\kappa}^\pm(t)$ are related to the parameters of the classical solution $\tilde{\kappa}^\pm (t)=  (\alpha^\pm(t))^{2/3} ,$ cf. Eq.~(\ref{rho0Near}). The density gradient (\ref{deltaXrhoIsAi2}) exhibits pronounced oscillations, ``quantum ripples'' in the vicinity of the classical shock points, see the inset in Fig. \ref{typicalRho}. At times $t\gtrsim t_S$, the distance between the shock points is $x_+(t)-x_-(t)\sim\Delta x$. The characteristic length scale for the ripples $1/\tilde{\kappa}^\pm (t)\sim \Delta x/(\Delta N)^{2/3}$ is parametrically smaller. Making use of the identity ${\rm Ai}''(x)  =  x {\rm A_i}(x)$, one may integrate Eq.~(\ref{deltaXrhoIsAi2}) to find 
\begin{align}\label{LeadingRho}
&\delta\rho^{\pm }(x,t) \\
&\simeq {\tilde{\kappa}}^\pm(t)\!\left\{\left[{\rm Ai}'(\tilde{\kappa}^\pm (t)\delta x_\pm)\right]^2\!\! -\! (\tilde{\kappa}^\pm(t) \delta x_\pm) {\rm Ai}^2(\tilde{\kappa}^\pm  (t)\delta x_\pm ) \right\}. \nonumber
\end{align}
Equation (\ref{LeadingRho}) together with the derivative (\ref{deltaXrhoIsAi2}) is the main result of this Letter. A typical plot of $\rho(x)$ is shown in Fig. \ref{typicalRho}, with the inset showing the graph of $\partial_x \delta\rho^\pm(x)$ around  a branch meeting point.

To conclude, we found quantum corrections to the classical shocks forming in the course of evolution of a one-dimensional Fermi gas. The leading quantum effect is the formation of density ripples in the vicinity of the shock points. The quasiclassical parameter controlling the spatial structure and strength of the ripples scales as $(\Delta N)^{-2/3}$ with the excess number of particles $\Delta N$ in the evolving nonlinear wave.

We gratefully acknowledge collaboration with Adilet Imambekov (LG) and P. Wiegmann (EB), which have lead us to the current study. We thank A.~Abanov, A.~Mirlin, and V.~Protopopov for stimulating discussions, and the authors of \cite{Mirlin:Gutman:Density:Oscillations} for alerting us to their paper while we were finalizing the text of our work. The work was supported by Israel Science Foundation, Grant No. 852/11 (EB), Binational Science Foundation, Grants No. 2010345 (EB) and No. 2010366 (LG), and by NSF DMR Grant No. 1206612 (LG).

\bibliographystyle{apsrev4-1}
\bibliography{mybib}

\end{document}